\documentstyle[aps,prl,floats,graphicx,epsfig]{revtex}
\begin{document}
\draft

\title{Monopole, half-quantum vortices and nexus in chiral superfluids and
superconductors. }
\author{G.E. Volovik\\
Helsinki University of Technology,
Low Temperature Laboratory,\\
P.O.Box 2200,
FIN-02015 HUT,  Finland, \\
 Landau Institute for Theoretical Physics,
117334 Moscow,
Russia }

\date{\today}

\maketitle

\begin{abstract}
Two exotic objects  are still not identified experimentally in chiral
superfluids
and superconductors. These are the half-quantum vortex, which plays the
part of the Alice string in relativistic theories \cite{Schwarz}, and the
hedgehog
in the
$\hat{\bf l}$ field, which is the counterpart of the Dirac magnetic monopole.
These two objects of different dimensionality are topologically  connected.
They
form the combined object which is called nexus in relativistic theories
\cite{Cornwall}. Such combination will allow us to observe half-quantum
vortices
and monopoles in several realistic geometries.
\end{abstract}

In relativistic quantum fields nexus is the monopole, in which
$N$ vortices of the group $Z_N$ meet at a center (nexus) provided the total
flux of
vortices adds to zero (mod $N$)
\cite{Cornwall,Reinhardt}. In a chiral superfluid with the order parameter
of the $^3$He-A type, the analog of the nexus is the hedgehog in the
$\hat{\bf l}$
field, in which 4 vortices meet, each with the circulation quantum number
$N=1/2$.
The total topological charge of the four vortices is $N=2$ which is
equivalent to
$N=0$ because the homotopy group, which describes the
$^3$He-A vortices, is $\pi_1=Z_4$ \cite{VolMinTop}, and thus $N=0$ (mod 2).
Each
$N=1/2$  vortex is the $1/4$ fraction of the "Dirac string" in  $^3$He-A, the
latter is the $N=2$ vortex terminating on the hedgehog
\cite{Blaha,VolMin}. The hedgehog in the
$\hat{\bf l}$ field plays a part of the Dirac magnetic monopole: The
distribution of the vector potential of the
electromagnetic field ${\bf A}$ in the vicinity of the hedgehog in
the electrically charged version of the $^3$He-A (the chiral $p$-wave
superconductor) is similar to that in the vicinity
of magnetic monopole (see e.g. \cite{MonopoleAPhase}).

The order parameter describing the  vacuum manifold in chiral $p$-wave
superfluid/superconductor ($^3$He-A and also possibly the layered
superconductor Sr$_2$RuO$_4$~\cite{Rice}) is
\begin{equation}
A_{\alpha i}= \Delta \hat d_\alpha( \hat e^{(1)}_i + i\hat e^{(2)}_i) ~.
\label{OrderParameter}
\end{equation}
Here $\hat{\bf d}$ is the unit vector of the spin-space anisotropy;  $\hat{\bf
e}^{(1)}$ and $\hat{\bf
e}^{(2)}$ are unit mutually orthogonal vectors in the orbital space, they
determine the superfluid velocity of the chiral condensate ${\bf
v}_s={\hbar\over
2m}\hat e^{(1)}_i\nabla \hat e^{(2)}_i$, where $2m$ is the mass of the Cooper
pair; the orbital momentum vector is $\hat{\bf l}=\hat{\bf e}^{(1)}\times
\hat{\bf
e}^{(2)}$. The half-quantum vortex results from the identification of the
points
$\hat{\bf d}$ ,  $\hat{\bf e}^{(1)}+i\hat{\bf e}^{(2)}$ and $-\hat{\bf d}$ ,
$-(\hat{\bf e}^{(1)}+i\hat{\bf
e}^{(2)})$, which correspond to the same order parameter
Eq.(\ref{OrderParameter}). It is the combination of the $\pi$-vortex and
$\pi$-disclination in the $\hat{\bf d}$ field:
\begin{equation}
\hat{\bf d}=\hat{\bf x} \cos{\phi\over 2} +\hat{\bf y} \sin{\phi\over 2} ~,~
\hat{\bf e}^{(1)}+i\hat{\bf e}^{(2)}=e^{i\phi/2}(\hat{\bf x} +i\hat{\bf y})~,
\label{HalfQuantumVortex}
\end{equation}
where $\phi$ is the azimuthal angle around the string.

The hedgehog in the orbital momentum field, $\hat{\bf l}=\hat{\bf r}$, produces
the  superfluid velocity field  (or the vector potential in the corresponding
superconductor):
\begin{equation}
{\bf v}_s={e\over mc}   {\bf A}~,~  {\bf A}=  \sum_a  {\bf
A}^a~,
\label{SuperfluidVelocity}
\end{equation}
where ${\bf A}^a$ is the vector potential for the Dirac monopole with the
$a$-th
Dirac string, $N_a$ is the topological charge (number of circulation quanta) of
the
$a$-th string.  Choosing the spherical coordinate system $(r,\theta,\phi)$
in such
a way that   the string
$a$ occupies the lower half-axis $z<0$, the vector potential ${\bf A}^a$ of
such
string can be written as \cite{MonopoleAPhase}:
\begin{equation}
{\bf A}^a={\hbar c\over 4e r} N_a \hat{\bf\phi} {1-\cos\theta\over
\sin\theta}~,
\label{DiracMonopoleField}
\end{equation}
The superfluid vorticity and the   corresponding magnetic field in
superconductor are
\begin{eqnarray}
\nabla\times {\bf v}_s=- {\hbar\over 4m}{{\bf r}\over r^3}\sum_aN_a +
{h\over 2m}\sum_a
N_a \int_0^R  dr~ \delta({\bf r}-{\bf r}_a(r))~,
\label{Vorticity}\\
 {\bf B}=-{\hbar c\over 4e}  {{\bf r}\over
r^3}\sum_aN_a + {hc\over 2}\sum_a
N_a \int_0^R  dr~ \delta({\bf r}-{\bf r}_a(r))~,~\sum_a
N_a=-2~.
\label{MagneticField}
\end{eqnarray}
Here ${\bf r}_a(r)$ is the position of the $a$-th line, assuming that the
lines are
emanating radially from the monopole, i.e. the coordinate along the line is the
radial coordinate. The regular part of the magnetic field  corresponds to the
monopole   with the magnetic charge
$g=\hbar c/2e$, the magnetic flux $4\pi g$ of the monopole is supplied  by the
Abrikosov vortices.  The lowest energy of the monopole occurs when all the
vortices emanating from the monopole have the lowest circulation number: this
means that there must be four vortices with
$N_1=N_2=N_3=N_4=-1/2$.

The half-quantum vortices are accompanied by the spin
disclinations. Assuming that the $\hat{\bf d}$-field is confined in the plane
the disclinations can be characterized by the winding numbers
$\nu_a$ which have values $\pm 1/2$ in half-quantum vortices. The corresponding
spin-superfluid velocity
${\bf v}_{sp}$  is
\begin{equation}
{\bf v}_{sp}={e\over mc}\sum_{a=1}^4 \nu_a {\bf A}^a~,~\sum_{a=1}^4 \nu_a=0~,
\label{SpinSuperfluidVelocity}
\end{equation}
where the last condition means the absence of the monopole in the spin
sector of
the order parameter. Thus we have $\nu_1=\nu_2=-\nu_3=-\nu_4= 1/2$.

The spin-orbit coupling can be neglected if the size of the bubble is less than
spin-orbit length (about $10~\mu m$ in $^3$He-A).
Assuming that the superfluid velocity is everywhere perpendicular to  $\hat{\bf
l}$ and has a form ${\bf v}_s=\tilde {\bf v}_s(\theta,\phi)/r$, the energy
of the
nexus in the spherical bubble of radius $R$ is
\begin{eqnarray}
E= \int_0^R  r^2dr \int d\Omega \left( {1\over 2} \rho_s {\bf v}_s^2 +
{1\over 2}
\rho_{sp} {\bf v}_{sp}^2  \right)= R   \int d\Omega \left( {1\over 2} \rho_s
\tilde{\bf v}_s^2 + {1\over 2}
\rho_{sp} \tilde{\bf v}_{sp}^2  \right)=\\
{1\over 2} R   \int d\Omega \left(  (\rho_s+\rho_{sp})\left[(\tilde{\bf
A}^1+ \tilde{\bf A}^2)^2 +(\tilde{\bf
A}^3+ \tilde{\bf A}^4)^2\right] + 2 (\rho_s-\rho_{sp}) (\tilde{\bf
A}^1+ \tilde{\bf A}^2) (\tilde{\bf
A}^3+ \tilde{\bf A}^4)\right)~, ~\tilde{\bf A}^a(\theta,\phi) ={mcr\over e}
{\bf
A}^a
 ~.
\label{Energy}
\end{eqnarray}

In the simplest case, which occurs in the ideal Fermi
gas approximation when the Fermi liquid corrections are neglected, one has
$\rho_s
=\rho_{sp} $ \cite{VolWol}. In this case   the
$1/2$-vortices with positive spin current circulation $\nu$ do not interact
with
$1/2$-vortices with negative $\nu$.  The energy minimum occurs when the
orientations of two positive-$\nu$ vortices are opposite, so that these two
${1\over 4}$ fractions of the Dirac strings form one line along the
diameter (see
Fig.1). The same happens for the other fractions with negative $\nu$. The
mutual
orientations of the two diameters is arbitrary in this limit. However, in
real $^3$He-A one has $\rho_{sp}<\rho_s $ \cite{VolWol}. If
$\rho_{sp} $ is slightly smaller than
$\rho_s$, the positive-$\nu$ and negative-$\nu$ strings repel each other,
so that
the equilibrium angle between them is $\pi/2$. In the extreme case
$\rho_{sp}\ll \rho_s $, the ends of four half-quantum vortices form
vertices of a
regular tetrahedron.

\begin{figure}[!!!t]
\begin{center}
\leavevmode
\epsfig{file=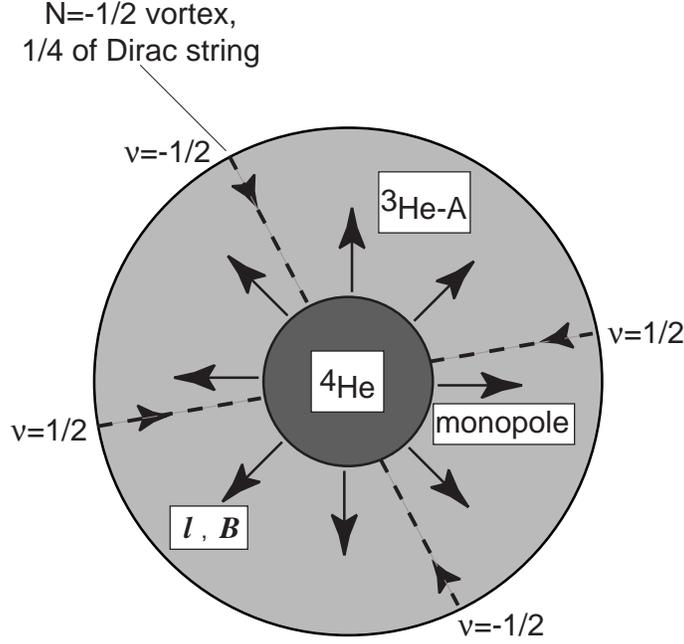,width=0.5\linewidth}
\caption[nexus]
    {Arrows outward show distribution of the orbital momentum ${\hat{\bf l}}$
field and simultaneously the distribution of superfluid vorticity $\nabla\times
{\bf v}_s$ in superfluid $^3$He-A or of magnetic field ${\bf B}$ in
a chiral superconductor. The  arrows inward show the direction of the
vorticity or
magnetic flux concentrated in 4 half-quantum vortices (dashed lines). The
charge
$\nu=\pm 1/2$ is the number of the circulation quanta of the spin current
velocity
${\bf v}_{sp}$. The stability of the monopole in the center of the droplet is
supported by the cluster of the
$^4$He liquid, which provides the radial boundary condition for the ${\hat{\bf
l}}$-vector. The cluster forms the core of the monopole.}
\label{nexus}
\end{center}
\end{figure}

Such monopole can be experimentally realized in the mixed $^4$He/$^3$He
droplets
obtained via the nozzle beam expansion of the He gases \cite{Vilesov}.
The $^4$He component of the mixture forms the cluster in a central region
of the droplet \cite{Vilesov}. If the size of the cluster is comparable
with the
size of the droplet, the radial distribution of the $\hat{\bf l}$ vector is
stabilized by the boundary conditions on the the surface of the droplet and
on the
boundary of the cluster (see Fig.1). The $^4$He cluster plays the part of
the core
of the nexus. The half-quantum vortices emanating from the nexus are well
defined
if the radius of the droplet exceeds the coherence length $\xi \sim
200-500~ \AA$.

In a $p$-wave superconductor such monopole will be formed in
a thin spherical layer. In Sr$_2$RuO$_4$ superconductor the spin-orbit coupling
between the spin vector $\hat{\bf d}$ and crystal lattice seems to align the
$\hat{\bf d}$ vector  along $\hat{\bf l}$\cite{Sigrist}. In this case the half
quantum vortices are energetically unfavourable, and instead of 4 half-quantum
vortices one would have 2 singly quantized vortices in the spherical shell.

Monopole of this kind can be formed also in the so called ferromagnetic Bose
condensate in optical traps. Such condensate is described by vector or spinor
chiral order parameter \cite{Ho}.

There are interesting properties of the system related to the fermionic
spectrum of
such objects. In particular, the number of fermion zero modes on $N=1/2$ vortex
under discussion is twice less than that on the vortex with $N=1$. This is
because such $N=1/2$ vortex can be represented as the $N=1$ vortex in one spin
component with no vortices in another spin component. Thus, according to
\cite{Volovik99}, in the core of the $N=1/2$ vortex there is one fermionic
level
(per 2D layer)  with exactly zero energy. Since the zero-energy level can be
either filled or empty, there is an entropy $(1/2) \ln 2$ per layer related
to the
vortex. The factor
$(1/2)$ appears because the particle  excitation coincides with the
antiparticle
(hole) excitation in superconductors, i.e. the quasiparticle is a
Majorana fermion, see also \cite{Read}. Such fractional entropy also arises
in the
Kondo problem\cite{Rozhkov}.  According to \cite{Ivanov}, the $N=1$ vortex
has spin
$S=1/4$ per layer, this implies the spin $S=1/8$ per layer for $N=1/2$ vortex.
Similarly the anomalous fractional charge of the $N=1/2$ vortex is 1/2 of that
discussed for the $N=1$ vortex \cite{Goryo}.

I thank M. Feigel'man and D. Ivanov for  discussions. This work was
supported in part by the
Russian Foundations for Fundamental Research
and by
European Science Foundation.

\end{document}